\documentclass[conference]{IEEEtran}
\IEEEoverridecommandlockouts
\usepackage{cite}
\usepackage{amsmath,amssymb,amsfonts}
\usepackage{algorithm}
\usepackage{algorithmic}
\usepackage{graphicx}
\usepackage{textcomp}
\usepackage{xcolor}
\usepackage{makecell}
\usepackage{caption}

\usepackage[inkscapelatex=false]{ svg}
\def\BibTeX{{\rm B\kern-.05em{\sc i\kern-.025em b}\kern-.08em
    T\kern-.1667em\lower.7ex\hbox{E}\kern-.125emX}}
\begin{document}

\title{WindTunnel - A Framework for Community Aware Sampling of Large Corpora}

\author{
\IEEEauthorblockN{Michael Iannelli}
\IEEEauthorblockA{
    \textit{Yext, Inc.} \\
    New York, USA \\
    miannelli@yext.com\\
    0000-0002-5967-2026}
}

\maketitle

\begin{abstract}
Conducting comprehensive information retrieval experiments, such as in search or retrieval augmented generation, often comes with high computational costs. This is because evaluating a retrieval algorithm requires indexing the entire corpus, which is significantly larger than the set of (query, result) pairs under evaluation. This issue is especially pronounced in big data and neural retrieval, where indexing becomes increasingly time-consuming and complex.

In this paper, we present WindTunnel, a novel framework developed at Yext to generate representative samples of large corpora, enabling efficient end-to-end information retrieval experiments. By preserving the community structure of the dataset, WindTunnel overcomes limitations in current sampling methods, providing more accurate evaluations.
\end{abstract}

\begin{IEEEkeywords}
neural information retrieval, retrieval augmented generation, graph mining, community structure
\end{IEEEkeywords}

\section{Introduction}

Yext offers a key product that leverages a knowledge graph to enhance entity search for enterprise clients. These entities can encompass documents, web pages, individuals, locations, and objects that a client may wish to access. The ground truth data used for search evaluation in this scenario is amassed from a diverse array of sources, including internal and external annotaters, feedback from end-users, and explicit rules formulated by clients.

\subsection{Motivation}
Although the volume of the ground truth data can be substantial, it is dwarfed by the size of the underlying corpora on which the search is executed. These corpora may comprise millions of primary and auxiliary entities, the latter being entities not designated for appearance in search results but instrumental in the indexing and retrieval processes.

\begin{figure*}
\begin{minipage}{0.58\textwidth}
\includegraphics[width=\linewidth]{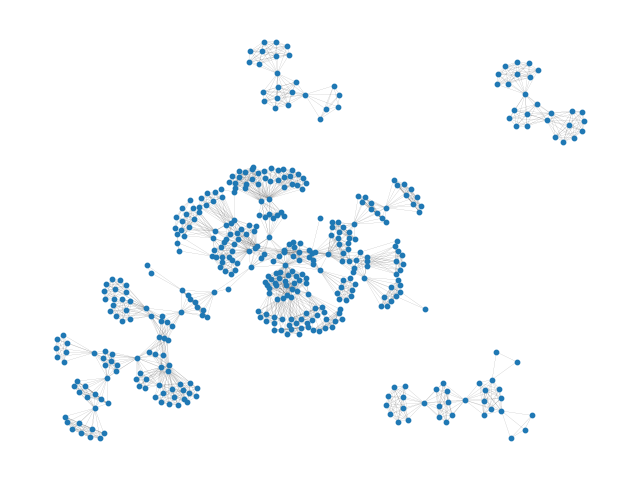}
\vspace*{-11mm}
\caption{A sample from the MSMarco corpus that preserves community structure, revealing the underlying community organization of documents. Each node in the graph represents a document, and an edge between nodes indicates that the corresponding documents share a common query.}
\label{fig_corpus_graph}
\end{minipage}
\hfill
\begin{minipage}{0.38\textwidth}
\includegraphics[width=\linewidth]{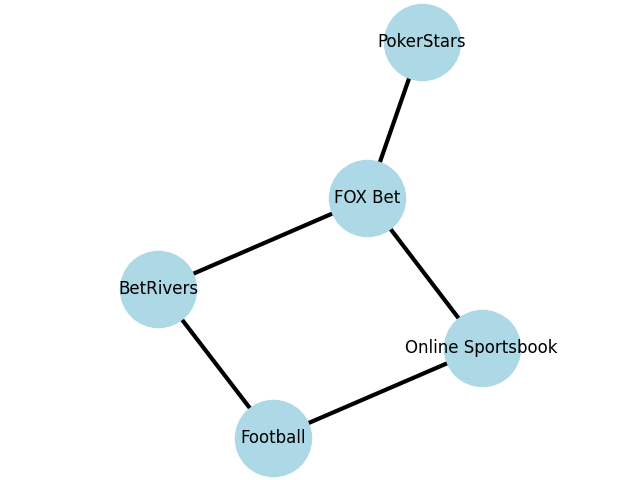}
\vspace*{10mm}
\caption{A detailed view of five nodes from a single community in the network shown in Figure \ref{fig_corpus_graph}, including document titles. Note the thematic consistency among document titles within the community; each document pertains to the subject of betting in some way.}
\label{fig_corpus}
\end{minipage}
\end{figure*}

Uniform random sampling is a common method for evaluating machine learning algorithms. However, it is ill-suited for information retrieval as it disrupts the latent community structure of the underlying data, leading to inflated precision metrics. Figure \ref{fig_corpus_graph} provides a visual representation of this issue in the context of neural information retrieval. Simplistically sampling entities with a uniform random probability amplifies the sparsity of entities in the embedded space, consequently impacting precision metrics.

\subsection{Community Structure in Corpora}
In this paper, we contend that the entities encompassed within typical information retrieval corpora exhibit a community structure. This arises inherently from the fact that these entities constitute a scale-free network, characterized by node degrees that adhere to a power-law distribution. The connections within this network can either be explicit, as in the case of hyperlinks linking one website to another \cite{Albert1999DiameterWeb}, or implicit, as in instances where two passages respond to the same query. We substantiate the latter assertion in Section \ref{powerlaw_section} by demonstrating that the node degrees in a widely recognized benchmark corpus conform to the Yule-Simon discrete power-law distribution.

\subsection{Further Challenges}
The issue of computationally intensive experimentation has been exacerbated by several contemporary phenomena:
\begin{itemize}
    \item Neural Information Retrieval: The need to rerun the embedding model over the entire corpus of entities each time there is a modification to the embedding model, making the indexing process particularly demanding.
    \item Large Language Models (LLM) and Retrieval Augmented Generation: Neural information retrieval serves as a crucial component of contemporary LLM applications, such as chatbots \cite{Lewis2020Retrieval-AugmentedTasks}.
    \item Big Data: The search space frequently encompasses millions or even billions of entities, and the necessity for re-indexing with each experiment increases linearly with the size of the corpus.
\end{itemize}

\subsection{Contributions}
The key contributions of this work are as follows:

\begin{itemize}
    \item We introduce WindTunnel, an innovative framework designed to construct a representative, smaller corpus from a larger one, thereby enabling more efficient experimentation.
    \item We present example implementations of WindTunnel Components using MapReduce Algorithms.
    \item Utilizing MSMarco as a benchmark corpus, we present evidence that the set of search results adheres to the Yule-Simon distribution, and consequently, exhibits community structure.
    \item Again, using MSMarco as a benchmark corpus, we offer experimental results that compare samples generated by WindTunnel to those obtained through uniform random sampling of passages.
\end{itemize}

\hfill
\section{Architecture and Implementation}

\begin{figure*}
\begin{center}
\includegraphics[width=0.8\textwidth]{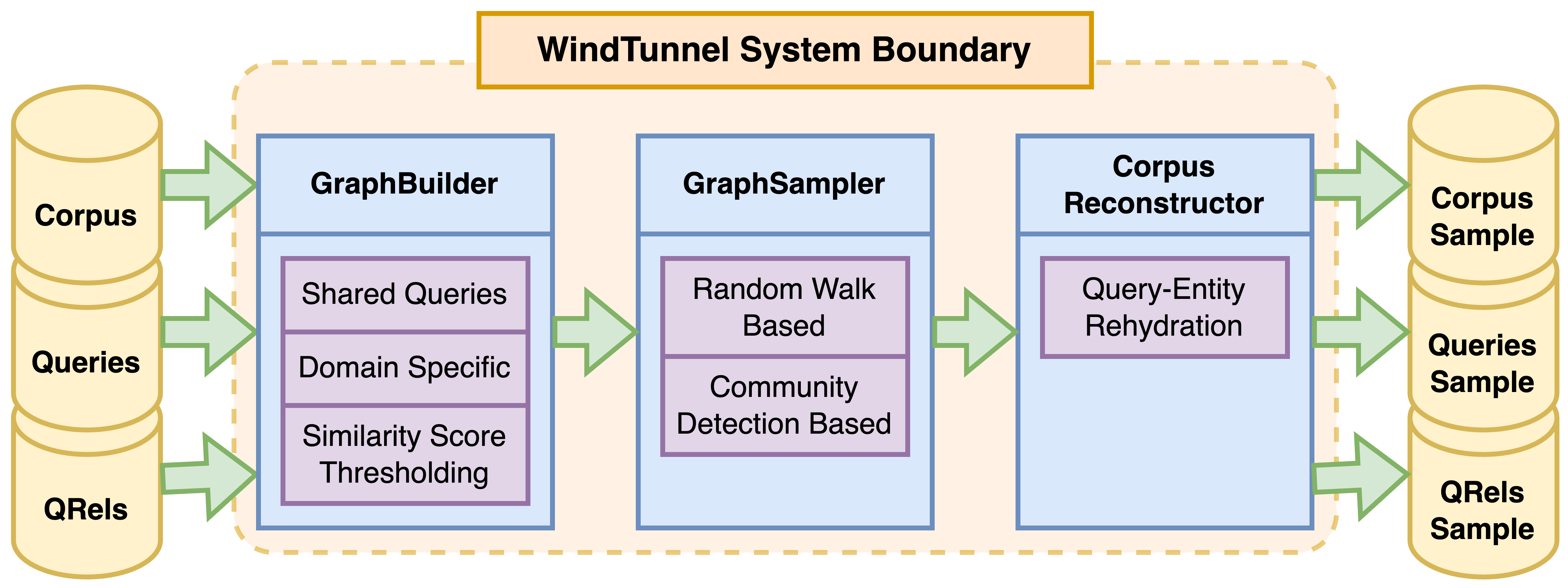}
\caption{High-Level Architecture of WindTunnel} \label{fig3}
\label{fig_architecture}
\end{center}
\end{figure*}

The WindTunnel Framework is composed of three main architectural components, as illustrated in Figure \ref{fig_architecture}. Designed to operate on extremely large datasets, WindTunnel is well-suited for implementation in distributed processing execution engines like Apache Spark \cite{Zaharia2010Spark:Sets}.

\begin{description}
    \item[\textbf{Input}:] \hfill \\
    The inputs to WindTunnel consist of three relational datasets:
    \begin{itemize}
        \item Queries: A table of benchmark queries with the schema $(query\_id, query\_content)$.
        \item Corpus: A table of entities with the schema $(entity\_id, entity\_content)$.
        \item Query Relevances (QRels): A table of relevance judgments that associate each query with entities in the corpus. The QRels table also includes a score indicating the relevance of the entity to the query.  The schema is $(entity\_id, query\_id, score)$.
    \end{itemize}
    \item[\textbf{GraphBuilder}:]  \hfill \\
    Generates relevant edges between entities in the target corpus. These edges originate from various sources, including:
    \begin{itemize}
        \item Shared queries, i.e., a single query is relevant to both entities.
        \item Domain-specific, i.e., one entity explicitly links to the other via hyperlink, shared ancestor, etc.
        \item Similarity score thresholding, i.e., the two entities have similar content, as estimated by a locality-sensitive hashing algorithm (LSH). See \cite{Halcrow2020Grale:Learning} for graph building methods using LSH.
    \end{itemize}    
    
    \item[\textbf{GraphSampler}:]  \hfill \\
    Implements a graph sampling process on the edges emitted by the GraphBuilder component. Several methods exist for sampling from a graph while preserving its essential properties \cite{Leskovec2006SamplingGraphs}.  For our experiments, we utilize the 2-phase, label-propagation - cluster sampling based approach described in Section \ref{graph_creation}.   
    \item[\textbf{CorpusReconstructor}:]  \hfill \\
    Reconstructs the corpus as a sample by taking the output edges from the GraphSampler and joining them with the relevant queries and entities from the original dataset.
    \item[\textbf{Output}:]  \hfill \\
    The output is a community-preserving sample of the input datasets with the same schema.
\end{description}

\section{Experiments}

For our experiments, we employ the MSMarco passages dataset \cite{Nguyen2016MSDataset}, a widely-used benchmark for information retrieval tasks. The dataset comprises a collection of query-passage rankings. To simplify our experimental setup, we consider only shared queries as the source of edges in our graph and select rankings with scores in the top 50\%.

\begin{figure*}
\begin{center}
\includegraphics[width=.8\textwidth]{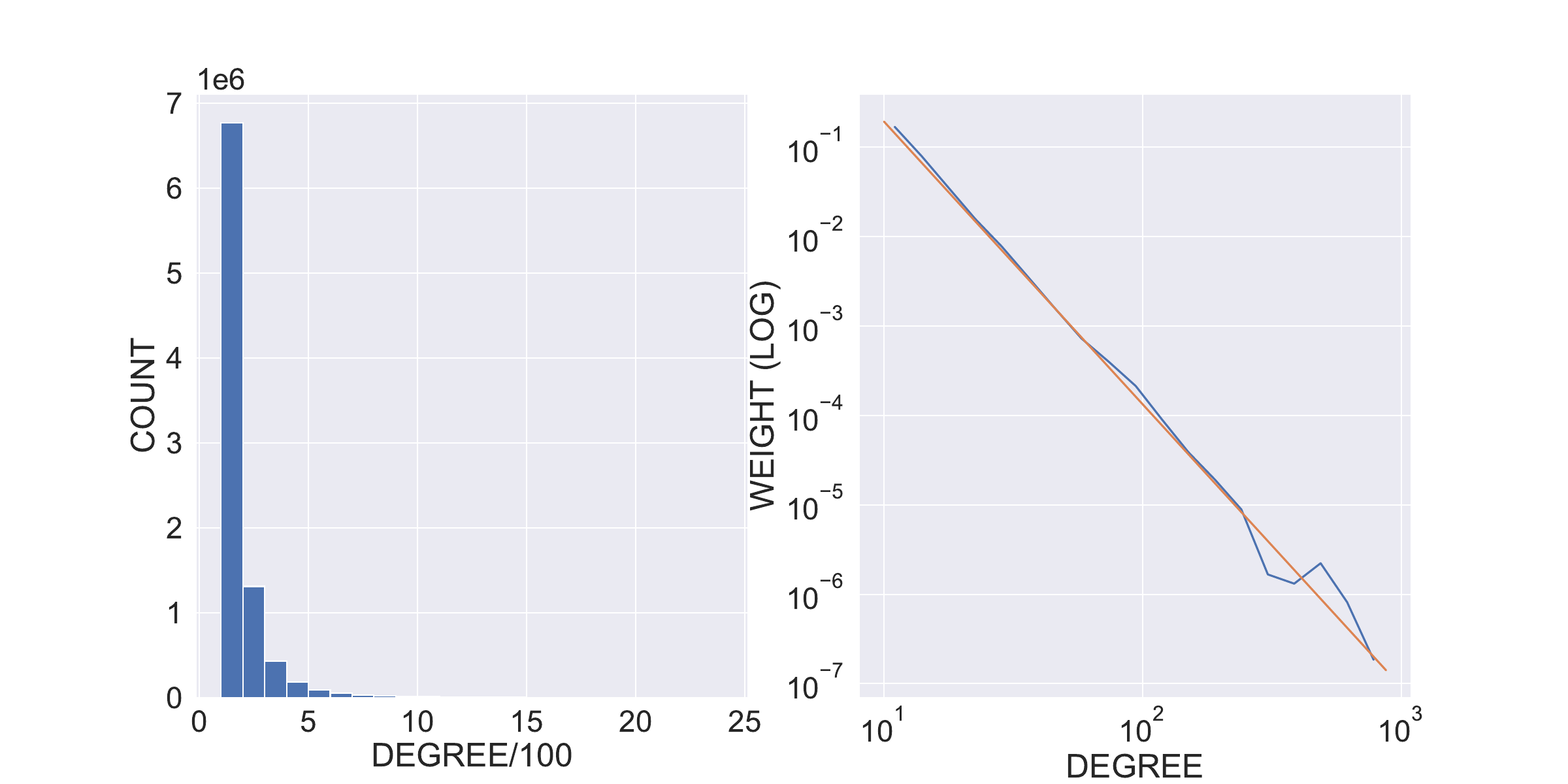}
\caption{Left: Histogram depicting the distribution of node degrees for MSMarco passages, where each node represents a passage, and two nodes are neighbors if their corresponding passages respond to the same query. The node degree of a node is the count of its neighbors. Right: A comparison of the MSMarco passages' node degree distribution (in blue) with a theoretical power-law distribution (in orange).}
\label{fig_degree_dist}
\end{center}
\end{figure*}

\subsection{Community Structure of Search Results}
\label{powerlaw_section}

We examine the degree distribution of passages within the MSMarco dataset. Figure \ref{fig_degree_dist} displays this distribution and contrasts it with a theoretical power-law distribution. A Yule-Simon distribution is fitted to the data using the Expectation-Maximization (EM) algorithm, as outlined in \cite{roberts2017AnModels}. We obtain a $\gamma$ value of 2.94 with a standard error of $8.4e-11$, which closely aligns with the expected value of 3, thereby confirming that the underlying data follows a power-law distribution \cite{Bollobas2001TheProcess}.

\subsection{End-to-End Semantic Search Experiments}

\begin{figure*}
\includegraphics[width=\textwidth]{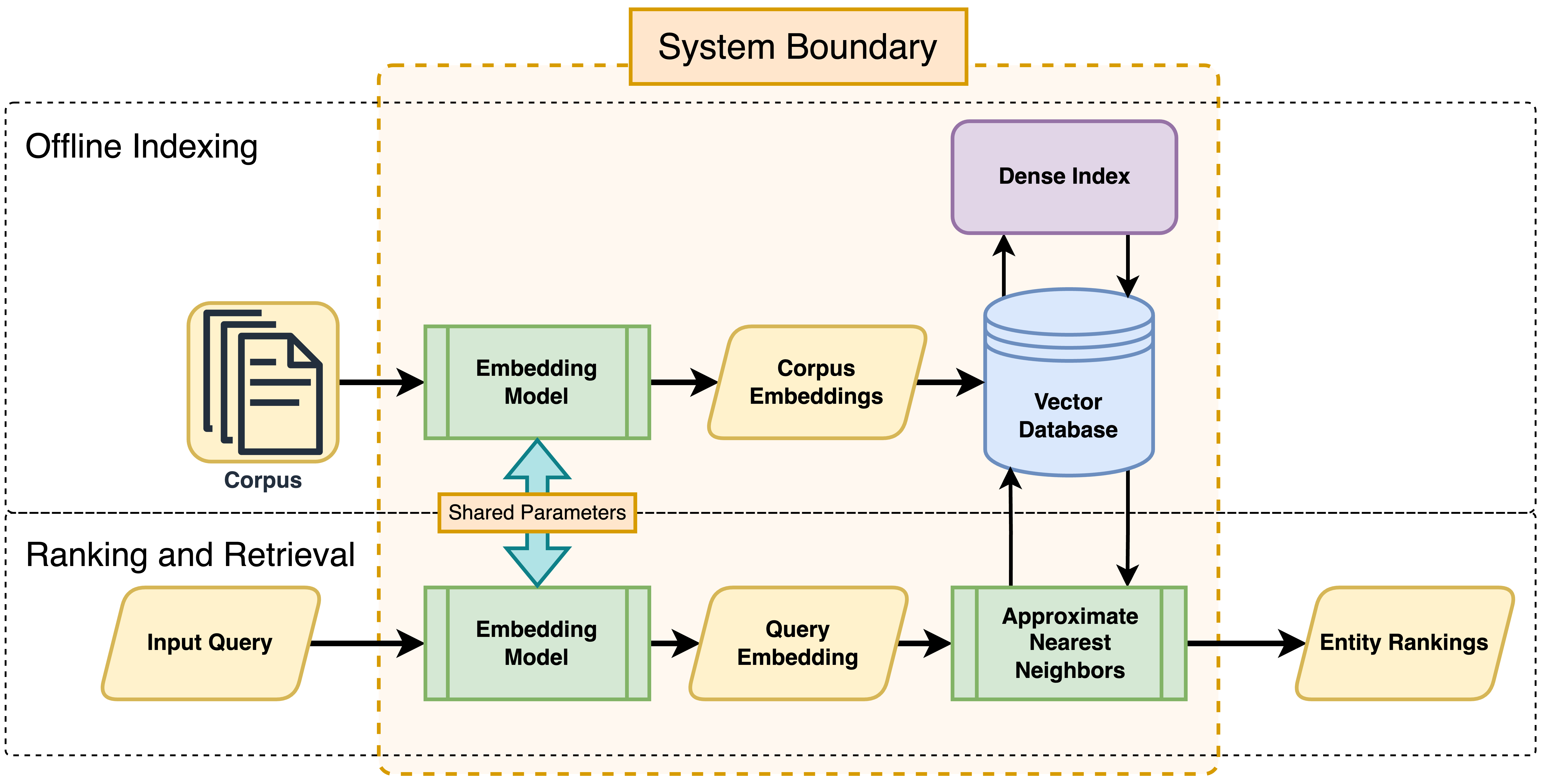}
\caption{Architecture of the Semantic Search Pipeline used for our experiments.  The pipeline consists of two high-level components operating asynchronously.  An offline indexing component accepts the input corpus, vectorizes the elements of that corpus using an embedding model, and indexes them into a vector database.  The online ranking and retrieval component accepts an input query from the end-user, vectorizes the query using the same embedding model from the offline indexing component, and then searches for approximate neighbors of that query vector in the vector database.} 
\label{ss_pipeline}
\end{figure*}

To evaluate our datasets, we assemble the semantic search pipeline depicted in Figure \ref{ss_pipeline}. The pipeline consists of:
\begin{itemize}
    \item An embedding model that projects passages and queries into a semantically meaningful vector space. We utilize a fine-tuned version of the transformer-based MPNet model \cite{Song2020MPNet:Understanding} for this task.
    \item A vector database capable of constructing dense vector indices, such as HNSW \cite{Malkov2018EfficientGraphs} or LSH \cite{Gionis1999SimilarityHashing}. For our experiments, we use the ivfflat index from the pgvector extension \cite{Yang2020PASE:Extension}.
    \item An Approximate Nearest Neighbors module responsible for retrieving and ranking entities based on their similarity to the input query. This functionality is generally integrated into the vector database.
\end{itemize}

We run three datasets through the pipeline: the complete MSMarco passage corpus, a uniform random sample of passages, and a WindTunnel-generated sample of passages. We measure Precision at 3 for each dataset, and the results are presented in Table \ref{end2end_results}.

\section{Results and Discussion}

Sampling passages uniformly and including associated queries leads to inflated precision metrics. This occurs because irrelevant neighboring nodes serve as distractors for the neural retrieval model, appearing in search results where they are not pertinent. This disruption of neighborhoods in the underlying dataset is a consequence of the sampling technique employed.

The precision of the model on the WindTunnel-generated sample was also higher than on the full corpus, albeit to a lesser extent. This discrepancy could arise from several factors. For instance, we only considered shared queries as a potential source of edges, and this single type of relation may not have been sufficient to capture the entire community structure of the dataset. Additional experiments are necessary to isolate these sources of discrepancy and fully understand their impact.

Table \ref{end2end_results} summarizes the mean precision@3 (p@3) scores for different sample types. The p@3 score reflects the relevance percentage of entities responding to each query, averaged over approximately 10K queries in the sample. The WindTunnel sample achieved a higher p@3 score compared to the baseline (0.288 vs. 0.105), but uniform random sampling resulted in an even higher score (0.916). This suggests that WindTunnel's sampling method preserves relevance but introduces fewer distracting neighbors compared to uniform sampling.

\begin{table}
\centering
\caption{Mean precision@3 (p@3) score}
\label{end2end_results}
\begin{tabular}{|l|l|}
\hline
Sample Type &  p@3\\
\hline
Baseline (full corpus) & 0.105 \\
Uniform Random Sample & 0.916 \\
WindTunnel Sample (100K Passages) & 0.288 \\
\hline
\end{tabular}
\end{table}

The WindTunnel-generated dataset also results in more compact communities of passages, as shown in Table \ref{c2q_ratio}. This leads to a higher query density ($\rho_q$), meaning that the same passages are relevant to multiple queries. Consequently, a higher percentage of passages in the dataset are returned for each query. For a sample of 30k queries, WindTunnel's query density was nearly three times that of the uniform random sample (0.294 vs. 0.106).

\begin{table}
\centering
\caption{Query Density ($\rho_q$) for a sample of 30k queries}
\label{c2q_ratio}
\begin{tabular}{|l|l|}
\hline
Sample Type &  $\rho_q$\\
\hline
Uniform Random Sample & 0.106\\
\makecell[l]{WindTunnel Sample (100K Passages)} & 0.294\\
\hline
\end{tabular}
\end{table}

The compactness of the WindTunnel sample's passage communities allows the retrieval model to focus on more densely connected regions of the corpus, improving the likelihood of relevant passages being retrieved across multiple queries. These findings indicate that the WindTunnel sampling method balances between maintaining query relevance and minimizing distraction, offering an improvement over the baseline and uniform sampling methods.

\bibliography{references}
\bibliographystyle{plain}


\appendix
\subsection{Algorithms}
The WindTunnel framework is versatile, supporting various algorithms. This paper demonstrates specific implementations of the GraphBuilder and GraphSampler components that translate the framework's conceptual design into scalable, practical operations utilizing the MapReduce paradigm.

\subsubsection{Entity Graph Creation}
\label{graph_creation}
Our GraphBuilder component constructs a homogeneous entity affinity graph using shared queries between entities to create edges, as described in Algorithm \ref{alg:entity_affinity}.

The process begins by filtering query-entity pairs based on relevance scores, where $qrel$ maps each pair to a score $S_{qrel}$. Pairs exceeding a threshold $\tau$ proceed to the next phase. Entity pairs sharing common queries are then identified, and the affinity score $S_{affinity}$ between two entities is defined as the minimum $S_{qrel}$ along the two-hop path ($id_{e1} \rightarrow id_q \rightarrow id_{e2}$). This results in a weighted graph where edges represent related entities.

In Step 2, duplicate edges are removed, ensuring that for multiple shared queries, only the pair with the highest affinity score is retained.

\begin{algorithm}[H]
\caption{MapReduce Algorithm for Entity Affinity Graph Construction}
\label{alg:entity_affinity}
\begin{algorithmic}
\raggedright
    \STATE \textbf{Define} $qrel: (id_q, id_{e}) \mapsto S_{qrel}$
    \STATE \textbf{Step 1: QRel Threshold Filtering and Entity Affinity Pair Generation}
    \STATE \textbf{Map Phase:}
    \STATE \quad Input: $(id_q, id_{e})$
    \STATE \quad Map Function: 
    \STATE \quad \quad $S_{qrel} = qrel(id_q, id_{e})$
    \STATE \quad \quad Emit $(id_q, (id_{e}, S_{qrel}))$ if $S_{qrel} > \tau$
    \STATE \textbf{Reduce Phase:}
    \STATE \quad Reduce Function: 
    \STATE \quad \quad $\forall \: (id_{e1}, id_{e2}) \, \exists \: id_q :$
    \STATE \quad \quad $(id_q, id_{e1}) \in \text{input keys} \: \land $
    \STATE \quad \quad $(id_q, id_{e2}) \in \text{input keys} \: \land $
    \STATE \quad \quad $id_{e1} < id_{e2} $,
    \STATE \quad \quad Compute $S_{affinity}$ as
    \STATE \quad \quad $\min(qrel(id_q, id_{e1}), qrel(id_q, id_{e2}))$
    \STATE \quad Output: $(id_{e1}, id_{e2}, S_{affinity})$
    \vspace{\baselineskip}

    \STATE \textbf{Step 2: Select Entity Affinity Pairs}
    \STATE \textbf{Map Phase:}
    \STATE \quad Input: $(id_{e1}, id_{e2}, S_{affinity})$
    \STATE \quad Map Function: 
    \STATE \quad \quad Emit $((id_{e1}, id_{e2}), S_{affinity})$
    \STATE \textbf{Reduce Phase:}
    \STATE \quad Reduce Function: 
    \STATE \quad \quad Compute $S_{affinity}^*$ as $max(\text{input values})$

    \STATE \quad Output: $(id_{e1}, id_{e2}, S_{affinity}^*)$
    
\end{algorithmic}
\end{algorithm}

\subsubsection{Graph Sampling}
Our implementation of the GraphSampler adopts a 4-step distributed algorithm.  Steps 1 to 3 identify related groups of entities using a label propagation inspired algorithm \cite{Raghavan2007NearNetworks} and Step 4 selects a subset of these groups via cluster sampling.  Pseudo-code for a MapReduce implementation is detailed in Algorithm \ref{alg:label_propagation_sampling}.

Steps 1 to 3 utilize a weighted label propagation based method to assign a label to each entity where the label indicates which community the entity belongs to.  The algorithm is initialized by setting each entity's label to its entity id.  The iteration phase of the algorithm aggregates over the labels of each entity's neighbors, computing the sum of affinities for each label.  The label with the max of affinity sums is taken and assigned to the node.  Since label propagation is not guaranteed to converge, the iteration is terminated after a set number of rounds.  Finally step 4 counts the number of entities assigned to each label and outputs each label in proportion to its total share of the entire corpus.

\begin{algorithm}[H]
\caption{MapReduce Algorithm for Label Propagation and Weighted Sampling}
\label{alg:label_propagation_sampling}
\begin{algorithmic}
\raggedright
    \STATE \textbf{Step 1: Instantiation}
    \STATE \textbf{Map Phase:}
    \STATE \quad Input: $(id_{e1}, id_{e2}, W)$
    \STATE \quad Map Function: 
    \STATE \quad \quad Emit $(id_{node}, id_{neighbor}, W, L):$ 
    \STATE \quad \quad \quad $id_{node} = id_{e1} \:\land $
    \STATE \quad \quad \quad $id_{neighbor} = id_{e2}\:\land$
    \STATE \quad \quad \quad $L = id_{e1}$
    \STATE \quad \quad Emit $(id_{node}, id_{neighbor}, W, L):$ 
    \STATE \quad \quad \quad $id_{node} = id_{e2} \:\land $
    \STATE \quad \quad \quad $id_{neighbor} = id_{e1}\:\land$
    \STATE \quad \quad \quad $L = id_{e2}$
    \STATE \textbf{Reduce Phase:}
    \STATE \quad Reduce Function: Pass-through
    \STATE \quad Output: $(id_{node}, id_{neighbor}, W, L)$
    \vspace{\baselineskip}

    \STATE \textbf{Step 2: Iteration}
    \STATE \textbf{Map Phase:}
    \STATE \quad Input: $(id_{node}, id_{neighbor}, W, L)$
    \STATE \quad Map Function: 
    \STATE \quad \quad Emit $(id_{node}, (id_{neighbor}, W, L))$
    \STATE \textbf{Reduce Phase:}
    \STATE \quad Reduce Function:
    \STATE \quad \quad Let $S(L) = \sum_{i : L_i = L} W_i$ 
    \STATE \quad \quad for each unique $L$
    \STATE \quad \quad $L^* = \underset{L}{\mathrm{argmax}}\ S(L)$
    \STATE \quad Output: $(id_{\text{node}}, id_{\text{neighbor}}, W, L^*)$
    \vspace{\baselineskip}
    
    \STATE \textbf{Step 3: Termination}
    \STATE \textbf{Map Phase:}
    \STATE \quad Input: $(id_{\text{node}}, id_{\text{neighbor}}, W, L)$
    \STATE \quad Map Function:
    \STATE \quad \quad Emit $(id_{node}, L): $
    \STATE \quad \quad \quad $id_{node} < id_{neighbor}$
    \STATE \textbf{Reduce Phase:}
    \STATE \quad Reduce Function: Pass-through
    \STATE \quad Output: $(id_{node}, L): $
    \vspace{\baselineskip}

    \STATE \textbf{Step 4: Sampling}
    \STATE \textbf{Define} $N:$ total number of entities 
    \STATE \textbf{Map Phase:}
    \STATE \quad Input: $(id_{node}, L): $
    \STATE \quad Map Function:
    \STATE \quad \quad Emit $(L, id_{node})$
    \STATE \textbf{Reduce Phase:}
    \STATE \quad Reduce Function:
    \STATE \quad \quad Emit $L$ with probability $\frac{|L|}{N}:$
    \STATE \quad \quad \quad$|L|$ = $count($input values$)$
    \STATE \quad Output: $L$

\end{algorithmic}
\end{algorithm}
\noindent \textbf{Note:} $W$ is the affinity score between entities, and $L$ is the label for the node ID.

\end{document}